\documentstyle[epsfig,12pt]{article}
\newcommand{\la}[1]{\label{#1}}
\newcommand{\ur}[1]{(\ref{#1})}

\newcommand{\eq}[1]{eq.~(\ref{#1})}

\newcommand{\Eq}[1]{Eq.~(\ref{#1})}

\newcommand{\e}{\epsilon}

\def\Tr{\mbox{Tr}}
\def\beq{\begin{equation}}
\def\eeq{\end{equation}}
\def\bea{\begin{eqnarray}}
\def\eea{\end{eqnarray}}

\begin{document}
\thispagestyle{empty}
\begin{flushright} NORDITA-2000/77 HE
\end{flushright}
\vskip 1.5true cm
\begin{center}
{\Large\bf Gauge-Invariant Formulation of the $d=3$ \\
\vskip .2true cm
Yang--Mills Theory}

\vskip 1true cm

{\large\bf Dmitri Diakonov$^{\diamond *}$ and Victor Petrov$^*$} \\
\vskip .8true cm
$^\diamond$ {\it NORDITA, Blegdamsvej 17, DK-2100 Copenhagen \O,
Denmark} \\
\vskip .5true cm
$^*$ {\it St.Petersburg Nuclear Physics Institute, Gatchina 188 350,
Russia} \\
\vskip .2true cm
\end{center}
\vskip .8true cm
\begin{abstract}
\noindent We write down the Yang--Mills partition function and the
average Wilson loop in terms of local gauge-invariant variables being
the six components of the metric tensor of dual space. The Wilson loop
becomes the trace of the parallel transporter in curved space, else
called the gravitational holonomy. We show that the external
coordinates mapping the $3d$ curved space into a flat $6d$ space
play the role of glueball fields, and there is a natural mechanism
for the mass gap generation.\\

\noindent PACS: 12.38.Aw, 12.38.Lg. Keywords: Wilson loop, glueballs, 
quantum gravity 
\end{abstract}

\vskip 1true cm

{\bf 1.} The problem of formulating Yang--Mills (YM) theory in terms of
gauge-invariant variables has been with us for the last 25 years.
Several approaches have been suggested to trade the YM potential for
some gauge-invariant variables. However, usually it is difficult to
incorporate matter in such approaches as it couples not to
gauge-invariant variables but to the gauge potential $A_\mu$.

The simplest but typical matter is the probe source, or the Wilson loop.
It is given by a path-ordered exponent of $A_\mu$, and it is difficult
to present it in terms of any gauge-invariant variables. At some point
it has been suggested that Wilson loops themselves are the only
reasonable gauge-invariant variables, and the theory should be
reformulated in terms of loop dynamics \cite{AMP,MM}. Because Wilson
loops are non-local objects there have been no decisive success on this
way, in spite of enormous efforts taken in twenty years.

In this Letter we suggest an approach which is opposite in
spirit to loop dynamics. Namely, we reformulate the YM theory in terms
of {\bf local} gauge-invariant variables, such that Wilson loops are
also expressible through these variables. We demonstrate it in a simple
case of $SU(2)$ theory in $d=3$ Euclidean dimensions. Four dimensions
and higher groups are, most probably, doable as well, and we comment on
that at the end of the paper.
\vskip .5true cm

{\bf 2.} The basis for our analysis is the reformulation of the
$SU(2)$ YM theory in $d=3$ in terms of quantum gravity. This has been
performed both in the continuum \cite{Lun1,AMS,DP4} by using the
first-order formalism and on the lattice \cite{ACSM,DP4} by making
duality transformation on the lattice and then passing to the continuum
limit. We start by briefly outlining the first derivation.

The YM partition function can be written with the help of an
additional Gaussian integration over the `dual field strength'
$G_i^a$ (a=1,2,3 are color and i=1,2,3 are Euclidean indices):
\bea
\nonumber
{\cal Z}&=& \int DA_i^a\, \exp\, -\frac{1}{4g_3^2}\int d^3x\,
F_{ij}^a(A)F_{ij}^a(A)\;\;\;\;\qquad \left[ F_{ij}^a(A)=
\partial_iA_j^a-\partial_jA_i^a+\e^{abc} A_i^bA_j^c\right] \\
\la{Z1}
&=&\int\!
DG_i^a\,DA_i^a\,\exp\int\!d^3x\left[-\frac{g^2_3}{2}G_i^aG_i^a
+\frac{i}{2}\e^{ijk}F_{ij}^a(A)G_k^a\right].
\eea

Gaussian integration over $A_i^a$ can be easily performed, the
saddle-point $\bar A_i^a$ being defined by the equation
\beq
\e^{ijk}(\partial_j\delta^{ab}+\e^{acb}\bar A_j^c)G_k^b
=\e^{ijk}\,D_j^{ab}(\bar A)G_k^b=0.
\la{sp}\eeq
This equation is generally satisfied if
\beq
D_j^{ab}(\bar A)G_k^b=\Gamma_{jk}^iG_i^a
\la{Gam}\eeq
where $\Gamma$ is symmetric in subscripts. \Eq{Gam} is known to be
satisfied in general relativity, provided one makes the following
identification:
\begin{itemize}
\item $G_i^a\equiv e_i^a$ where $e_i^a$ is a {\em dreibein} such that
the {\em metric tensor} is the standard
\beq
g_{ij}=e_i^ae_j^a,\qquad g^{ij}=e^{ai}e^{aj},
\qquad e_i^ae^{bi}=\delta^{ab},\qquad \det e_i^a=\sqrt{g};
\la{metric}\eeq
\item $\Gamma$ is the standard {\em Christoffel symbol},
\beq
\Gamma_{i,jk}=\frac{1}{2}\left(\partial_kg_{ik}+\partial_jg_{ik}
-\partial_ig_{jk}\right),\qquad \Gamma_{jk}^i=g^{il}\Gamma_{l,jk};
\la{Gam1}\eeq
\item $\bar A_i^c\equiv -\frac{1}{2}\e^{abc}\omega_i^{ab}$
where $\omega$ is the {\em spin connection},
\beq
\omega^{ab}_i =-\omega^{ba}_i
=\frac{1}{2}e^{ak}(\partial_i e_k^b-\partial_k e_i^b)
-\frac{1}{2}e^{bk}(\partial_i e_k^a -\partial_ke_i^a)
-\frac{1}{2}e^{ak}e^{bj}e^c_i (\partial_ke_j^c-\partial_je_k^c).
\la{sc}\eeq
\end{itemize}
It is easy to check that with the above definitions \eq{Gam} and
hence \eq{sp} are fulfilled. \Eq{sc} gives an explicit solution
for the saddle-point potential $\bar A_i^a$ through the dual field
strength $G_i^a=e^a_i$.

Pursuing further the analogy with Riemannian geometry one can define
the {\em covariant derivative} in curved space,
\beq
\left(\nabla_{\!i}\right)^k_l=\partial_i\delta^k_l+\Gamma^k_{il},
\la{covder}\eeq
whose commutator is the {\em Riemann tensor},
\bea
\la{R1}
[\nabla_i\nabla_j]^k_l=R^k_{\;lij}&=&
\partial_i\Gamma^k_{jl}-\partial_j\Gamma^k_{il}
+\Gamma^k_{im}\Gamma^m_{jl}-\Gamma^k_{jm}\Gamma^m_{il}\\
\la{R2}
&=&\e^{abc}F_{ij}^a(\bar A)e^{bk}e^c_l,\qquad e^c_l\equiv G^c_l,
\eea
related, as we see, to the YM field strength at the saddle point
$\bar A$. More directly, we get from \eq{R2} the YM field strength
at the saddle point expressed in terms of gravity quantities,
\beq
F_{ij}^a(\bar A)=\frac{1}{2}R^m_{\;lij}\,\e^{abc}\,e^b_me^{cl}.
\la{F1}\eeq
Therefore, the second term in \eq{Z1} is, at the saddle
point, nothing but the Einstein--Hilbert action:
\beq
\e^{ijk}F_{ij}^a(\bar A)G_k^a
=\sqrt{g}R,\qquad R\equiv R^m_{\;lmn}g^{ln},
\la{EH}\eeq
where we have used the identity
\beq
\e^{abc}e^a_ke^b_m=\sqrt{g}\,\e_{kmn}\,e^{cn}.
\la{id1}\eeq
The first term in \eq{Z1} is
\beq
G^a_iG^a_i=e^a_ie^a_i=g_{ii}.
\la{eth1}\eeq

Integrating over $A^a_i$ in \eq{Z1} gives a pre-exponential factor
$(\det G^a_i)^{-\frac{3}{2}}= g^{-\frac{3}{4}}$.  The
resulting integral in \eq{Z1} is over the 9 components of the dreibein
$G^a_i$ which can be rotated by the 3-parameter gauge transformation
$O^{ab}$. However, both the action and the measure depend on
$G^a_i(x)$ only through the metric tensor $g_{ij}$ which is
{\em gauge-invariant}. Writing
$d^{(9)}e^a_i=d^{(3)}O^{ab}\,d^{(6)}\!g_{ij}g^{-\frac{1}{2}}$
one can pass to integrating over the metric. One gets finally
the YM partition function \ur{Z1} identically rewritten in terms of 6
gauge-invariant variables \cite{AMS,DP4}:
\beq
{\cal Z}=\int Dg_{ij}\,g^{-\frac{5}{4}}
\exp\int\!d^3x\,\left[-\frac{g_3^2}{2}\, g_{ii}+\frac{i}{2}\sqrt{g}R(g)
\right].
\la{Z2}\eeq

The second term, the Einstein--Hilbert action, is
diffeomorphism-invariant and depends actually only on three variables.
The first term is not, and we call it the `ether' term \cite{DP4} as it
corresponds to an isotropic and homogeneously distributed `matter'
which, however, spoils the general covariance. It distinguishes
the YM theory from the pure gravity theory (which is a topological
field theory \cite{Wit}), and is responsible for the propagation of
transverse gluons in the perturbative regime \cite{DP4}. We note
that the integration measure also differs from the
diffeomorphism-invariant measure $\prod_{i\leq j}dg_{ij}g^{-2}$.

It is instructive to rewrite the partition function \ur{Z2} further
introducing 6 external coordinates
$w^\alpha(x),\quad \alpha=1,\ldots 6$, describing the embedding of
a general 3-dim Riemannian manifold into 6-dim flat space.
The metric induced by embedding is
\beq
g_{ij}=\partial_iw^\alpha\partial_jw^\alpha.
\la{m1}\eeq
Quantities $\Gamma^i_{jk},\,\sqrt{g}$ and $R$ written in terms of
$w^\alpha$ can be found in ref.\cite{DP4}.

The Jacobian for the change of integration variables from the
6 components $g_{ij}(x)$ to 6 external coordinates $w^\alpha(x)$
is the $6\times 6$ determinant made of second derivatives,
$Dg_{ij}=Dw^\alpha \det(w^\alpha_{ij})$. It is zero in the degenerate
case when the manifold can be embedded in less than 6 dimensions. It
terms of the external coordinates the YM partition function has the
following form (it is in this form the partition function has been
derived in ref.\cite{DP4} from making a duality transformation on the
lattice and then passing to the continuum limit):
\beq
{\cal Z}=\int Dw^\alpha(x)\,\det(\partial_i\partial_jw^\alpha)\,
g(w)^{-\frac{5}{4}}\,
\exp\int\!d^3x\,\left[-\frac{g_3^2}{2}\,
\partial_iw^\alpha\partial_iw^\alpha
+\frac{i}{2}\sqrt{g(w)}\;R(w)\right].
\la{Z3}\eeq
\vskip .5true cm

{\bf 3.} Let us now consider the average of the Wilson loop in
representation $J=\frac{1}{2},1,\frac{3}{2},\ldots$ along a closed
contour $x^i(\tau)$, defined as
\beq
W_J=
\frac{1}{2J+1}\Tr \;{\rm P}\; \exp\,
i\oint\!d\tau\,\frac{dx^i}{d\tau}\, A_i^a\,T^a
\la{WJ}\eeq
where $T^a$ are the $SU(2)$ generators in representation $J$. It is
not at all clear that one can rewrite it in terms of local
gauge-invariant variables $g_{ij}$ or $w^\alpha$. A hint that it may be
possible comes from the dual formulation on the lattice where the
Wilson loop appears to be a product of certain $9j$ symbols along the
contour \cite{DP4}, each of which is gauge-invariant. We shall show
that in the gravity formulation the average $W_J$ becomes the average
of a parallel transporter in curved space, else called the holonomy,
taken in the appropriate representation (spinor, vector, etc.)

To that end it is convenient to write down the Wilson loop in a form
where the path ordering is traded for an additional functional
integration over a unit 3-vector ${\bf n}$ `living' on a
surface spanned on the contour \cite{DPwl},
\bea
\la{wl1}
W_J&=&\int D{\bf n}(\sigma,\tau)\,
\exp\left[iJ\int d\tau \frac{dx^i}{d\tau}(A^a_i\,n^a)
+\frac{iJ}{2}\int d^2S^{ij}\,
\e^{abc} n^a\partial_i n^b\partial_j n^c\right]\\
\la{wl2}
&=&\int D{\bf n}(\sigma,\tau)\,
\exp\left[\frac{iJ}{2}\int d^2S^{ij}\left(-F_{ij}^a\,n^a
+\e^{abc}\,n^a\left(D_i n\right)^b\left(D_jn\right)^c\right)\right].
\eea
This representation for the Wilson loop is convenient because $A^a_i$
comes linearly in the exponent. The second line is a `non-Abelian
Stokes theorem'; it is written down in an explicitly gauge-invariant
form but actually the terms quadratic in $A_i^a$ cancel out.

One defines the average Wilson loop by inserting \eq{WJ} or,
equivalently, \eq{wl1} into the partition function \ur{Z1}. Since the
integral over $A_i^a$ is Gaussian and since the source term $A^a_in^a$
is linear in $A^a_i$, it is sufficient to replace $A_i^a$ in the Wilson
loop by the saddle-point value $\bar A^a_i$. One can disregard the
quadratic point-like self-interacting term emerging from this
replacement, as being independent of the dynamics.

Meanwhile, $\bar A^a_i$ is given by the spin connection, see \eq{sc}.
The path-ordered exponent of the spin connection determines the parallel
transporter of spinors in curved space. Parallel transporters of higher
spins can be always built from spinors. We, thus, come to the
conclusion that in the quantum-gravity formulation of the YM theory
one can calculate the Wilson loop as the trace of a parallel
transporter along a closed contour in curved space.

\vskip .5true cm
{\bf 4.}
Spin connection $\omega^{ab}_i$ is defined by the dreibein $e^a_i$
which is not uniquely determined by the metric tensor. In principle,
for a given metric one can use any dreibein to calculate the Wilson
loop, but this is not very satisfactory. Since Wilson loops are
gauge-invariant and so is the metric tensor one would expect that
parallel transporters are expressible through the metric tensor (and
its derivatives) only.

Let us first consider Wilson loops in integer-$J$ representations.
In this case the problem of expressing them through the metric tensor
is immediatelly solved by the observation \cite{DPnonab} that the
holonomy defined via the spin connection is equal to that defined via
Christoffel symbols. For example, for the $J=1$ (vector) representation
one has \cite{DPnonab}:
\beq
W_{J=1}=\frac{1}{3}\left[{\rm P}\;\exp\,
-\oint\!d\tau\,\frac{dx^i}{d\tau}\,\omega_i(e)\right]^{aa}
=\frac{1}{3}\left[{\rm P}\;\exp\,
-\oint\!d\tau\,\frac{dx^i}{d\tau}\,\Gamma_i(g)\right]^k_k.
\la{WGWYM}\eeq

For half-integer representations the problem is more difficult
as there is no way to define the parallel transporter of spinors
other than via the spin connection. Nevertheless, this problem has been
solved in ref.\cite{DPnonab} with the help of the `gravitational
non-Abelian Stokes theorem' similar to \eq{wl2}. It is only for parallel
transporters of spinors along a closed loop and only when written down
in a surface form that the spinor holonomy can be expressed through the
metric tensor and its derivatives. Actually, one can write a unified
formula for the holonomy for any spin $J$; it is given by
a functional integral over a covariantly unit vector ${\bf m}$
`living' on a surface spanned on the contour \cite{DPnonab}:
\bea
W_J&=&\int D{\bf m}\,\sqrt{g}\;\delta(m^im^j\,g_{ij}-1)
\nonumber\\
\nonumber\\
&\times&\exp\;i\frac{J}{2}\int\! dS_k\,\sqrt{g}\,
\left[\left(R\delta^k_p-2R^k_p\right)\,m^p+\e^{ijk}\,
\e_{pqr}\,m^p(\nabla_{\!i}\,m)^q(\nabla_{\!j}\,m)^r\right]
\la{gravWL1}\eea
where $R^k_p$ is the Ricci tensor, $R=R^k_k$ is the scalar curvature
and $\nabla_{\!i}$ is the covariant derivative in curved space
\ur{covder}. Despite its surface form, \eq{gravWL1} is independed
of the way one draws the surface.

\Eq{gravWL1} gives a general solution to the problem of
expressing the YM Wilson loop in terms of the metric tensor; it should
be inserted into the partition function \ur{Z2} for calculating the
average Wilson loop over the ensemble of YM configurations. If external
coordinates $w^\alpha(x)$ are used as dynamical variables instead of
$g_{ij}(x)$ one has to express $R^i_k$ and $\Gamma^i_{jk}$ through
$w^\alpha$ first, and plug them in \eq{Z3}. In both cases the average
Wilson loop is expressed through {\bf local} gauge-invariant variables,
be it $g_{ij}$ or $w^\alpha$.

\vskip .5true cm
{\bf 5.} So far we have presented exact results, and now we would like
to speculate on how the expected mass gap for glueballs and the area
behavior of the Wilson loop might be achieved. We also give an example
where \eq{gravWL1} is considerably simlified.

A characteristic feature of the partition function \ur{Z3} is the
purely imaginary Newton constant. It damps field configurations leading
to fast oscillations. A trivial way to get rid of those oscillations is
to put $R=0$, which leads to the perturbative regime \cite{DP4}. A less
trivial example of how to avoid fast oscillations is to have the
Einstein--Hilbert action quantized so that the oscillating factor
becomes $\exp(2\pi i n)=1$. This is achieved {\it e.g.} in de Sitter
$S^{3}$ spaces of constant (quantized) curvature $R=54\pi^2/n^2$.

Indeed, the de Sitter space with curvature $R$ can be
generally parametrized by 4 external coordinates $w^A(x),\;A=1,\ldots,4$,
restricted to the $S^{3}$ sphere of radius $\sqrt{6/R}$:
\beq
\sum_{A=1}^4 w^A(x)w^A(x)=\frac{6}{R}.
\la{S3}\eeq
The square root of the metric determinant becomes a determinant itself
being equal to
\beq
\sqrt{g}=\frac{1}{6}\sqrt{\frac{R}{6}}\e^{ijk}\,\e^{ABCD}\,
\partial_iw^A\,\partial_jw^B\,\partial_kw^C\,w^D,
\la{sqrtg1}\eeq
so that the Einstein--Hilbert term in \eq{Z3} is
\beq
\frac{i}{2}\int d^3x\sqrt{g}R=i\,6\pi^2\sqrt{\frac{6}{R}}
N_{{\rm wind}},
\la{EH1}\eeq
where $N_{{\rm wind}}$ is the properly normalized winding number of the
mapping $S^{3}\mapsto S^{3}$ realized by $w^A(x)$. Putting it to be an
integer $n$ and demanding that the l.h.s. of \eq{EH1} is $2\pi i n$
we arrive to the quantization condition written above for the scalar
curvature. Such values of $R$ are preferred because there are no
damping oscillations then.

The `ether' term is the kinetic energy term for four fields $w^A(x)$
subject to the constraint \ur{S3}, therefore we are left with
a low-energy $O(4)$ sigma model. The large-$N$ sigma models are known
to develop a mass gap; technically it is achieved by introducing a
Lagrange multiplier for the constraint and then finding a saddle-point
value of this multiplier, which is justified at large $N$. It is not
clear if one can consider 4 as being much much larger than 1 but this
scenario seems probable. That would give a mass gap to glueballs
which we identify with the $w^A$ field.

Whether this dynamical regime leads to an area law for the holonomy in
half-integer representations, is less clear. On one hand it is
intuitively a rather natural outcome: the constant curvature is a
gauge-invariant analog of a constant field strength usually leading to
the area law. On the other hand we have not been able to prove it
directly so far, and the question remains open.

We would like to remark that a constant-curvature background is
a gauge-invariant analog of the long-sought master field leading to
a nonzero gluon condensate.

Another interesting regime is a cylindric space $S^{2}\times {\cal R}$
also of constant curvature $R$ but with one eigenvalue of the Ricci
tensor being zero. Such spaces are generally parametrized by four
coordinates $w$ with only three of them subject to the constraint
\beq
\sum_{a=1}^3(w^a)^2=\frac{2}{R},\qquad w^4\;{\rm arbitrary}.
\la{constr2}\eeq
In this case we have managed to find a general expression for an
arbitrary Wilson loop. It is given by the $SU(2)$ character
whose argument is the winding number of $w^a(x)$ \cite{DPnonab}:
\beq
W_J=\frac{1}{2J+1}\sum_{m=-J}^J\exp\, im\Phi,\qquad
\Phi=\left(\frac{2}{R}\right)^{\frac{3}{2}}\,\frac{1}{2}
\int dS_k\;\e_{abc}\;\e^{ijk}\;
\partial_iw^a\,\partial_jw^b\,w^c,
\la{WLmon1}\eeq
where $dS_k$ is the dual element of the surface spanned on the contour.
We note that \eq{WLmon1} gives an example of the Wilson loop
being expressed through local gauge-invariant variables, though it is
less general than \eq{gravWL1}.

For cylindric spaces the Einstein--Hilbert action is a full derivative,
\beq
\frac{i}{2}\int d^3x\sqrt{g}R=\frac{i}{8}R^{\frac{3}{2}}
\int d^3x\,\partial_k\left[w^4\,\e^{ijk}\,
\e^{abc}\partial_i w^a\,\partial_j w^b\,w^c\right],
\la{EH2}\eeq
and can be reasonably set to be zero, therefore there are no
oscillations in the partition function, too. The $O(3)$ sigma model we
are left with in this case may also develop a mass gap for glueballs.
However, this regime most probably does not lead to the area law. The
reason is that for large areas inside the Wilson loop the flux $\Phi$
in \eq{WLmon1} becomes the winding number of the mapping $S^{2}\mapsto
S^{2}$ but with the coefficient $4\pi$, therefore non-trivial mappings
$w^a(x)$ are not `felt' even in half-integer representations. A more
careful analysis in the spirit of the $CP^{N-1}$ model shows that the
static potential has a logarithmic, not linear, behavior.

\vskip 0.5true cm
{\bf 6.} In conclusion, we have reformulated both the YM partition
function and the average Wilson loop in terms of local gauge-invariant
variables, being actually the metric of the dual space.
It looks as a promising starting point for studying two features
expected in YM theory: the mass gap for glueballs and the area law.
Generalization to higher gauge groups and to $d=4$ is possible
{\it e.g.} along the lines suggested in refs.\cite{HJLS,Lun2}.
In $d=4$ a formula similar to \eq{gravWL1} exists, expressing the
holonomy through the metric tensor \cite{DPnonab}.



\begin{thebibliography}{99}

\bibitem{AMP}
A.M. Polyakov, {\it Nucl. Phys.} {\bf B164} (1980) 171.

\bibitem{MM}
Yu.M. Makeenko and A.A. Migdal, {\it Phys. Lett.} {\bf B88} (1979) 135.

\bibitem{Lun1}
F.A. Lunev, {\it Phys. Lett.} {\bf B295} (1992) 99.

\bibitem{AMS}
R. Anishetty, P. Majumdar and H.S. Sharatachandra, {\it Phys. Lett.}
{\bf B478} (2000) 373.

\bibitem{DP4}
D. Diakonov and V. Petrov, {\it Yang--Mills Theory in Three Dimensions
as Quantum Gravity}, hep-th/9912268.

\bibitem{ACSM}
R. Anishety, S. Cheluvaraja, H.S. Sharatchandra and M. Mathur,
{\it Phys. Lett.} {\bf 314B} (1993) 387.

\bibitem{Wit}
E. Witten, {\it Nucl. Phys.} {\bf B311} (1988/89) 46.

\bibitem{DPwl}
D. Diakonov and V. Petrov, {\it Sov. Phys. JETP Lett.} {\bf 49} (1989)
284; {\it Phys. Lett.} {\bf B224} (1989) 131; hep-th/9606104;
hep-lat/0008004.

\bibitem{DPnonab}
D. Diakonov and V. Petrov, {\it Non-Abelian Stokes Theorems in
Yang--Mills and Gravity Theories}, hep-th/0008035.

\bibitem{HJLS}
P.E. Haagensen and K. Johnson, {\it Nucl. Phys.} {\bf B439} (1995) 597;
P.E. Haagensen, K. Johnson and C.S. Lam, {\it Nucl. Phys.} {\bf B477} 
(1996) 273; R. Schiappa, {\it Nucl. Phys.} {\bf B517} (1998) 462.

\bibitem{Lun2}
F.A.Lunev, {\it Mod. Phys. Lett.} {\bf A9} (1994) 2281;
{\it J. Math. Phys.} {\bf 37} (1996) 5351, hep-th/9503133.




\end{thebibliography}
\end{document}